# Second-harmonic generation from hyperbolic plasmonic nanorod metamaterial slab

*Giuseppe Marino\*, Paulina Segovia, Alexey V. Krasavin, Pavel Ginzburg†, Nicolas Olivier, Gregory A. Wurtz, Anatoly V. Zayats.*

Department of Physics, King's College London, Strand, London, United Kingdom

ABSTRACT: Hyperbolic plasmonic metamaterials provide numerous opportunities for designing unusual linear and nonlinear optical properties. We show that the modal overlap of fundamental and second-harmonic light in an anisotropic plasmonic metamaterial slab results in the broadband enhancement of radiated second-harmonic intensity by up to 2 orders of magnitudes for TM- and TE-polarized fundamental light compared to a smooth Au film under TM-polarised illumination. The results open up possibilities to design tuneable frequency-doubling metamaterial with the goal to overcome limitations associated with classical phase matching conditions in thick nonlinear crystals.

Second-harmonic generation (SHG) is used frequently in modern laser technologies and applications[1]. The main advantage of this frequency doubling technique is the possibility of light generation at wavelengths rarely accessible via typical radiative decay channels of a gain medium and metrology of ultrashort pulses[2]. Approaches based on nonlinear optical crystals are constrained by phase matching conditions: fundamental and second-harmonic waves are required to propagate in phase to obtain optimum nonlinear energy conversion[1]. The operational bandwidth for ultrashort pulses is another limitation for frequency conversion in bulk nonlinear crystals[3].

To overcome these limitations and increase integrability, various nonlinear nanostructured media have been proposed based on nanoparticles, optical nanoantennas, and metamaterials[4-7]. These approaches use the light confinement and enhancement in plasmonic (metallic) nanostructures to locally increase the fundamental field strength upon which SHG depends quadratically[2]. In multi-resonant structures, SHG effect can be further enhanced with the additional resonant scattering provided by the nanostructure at the SH frequency[4]. Thus, conventional phase matching conditions can be replaced with a modal overlap at the fundamental and second-harmonic (SH) frequencies in the nanostructures[4-7]. In order to realise these conditions experimentally, various geometries have been used such as a broadband optical nanoantenna[6], a composite nanoparticle made of a rod and a V-shaped nanoantenna[4], and split-ring resonator based metamaterials[7].

In another approach to achieve broadband and alignment-free phase-matching, the so-called ε-near-zero (ENZ) metamaterials were proposed which by their nature achieve phase-matching conditions due to the electrostatic nature of the second-harmonic field inside the medium[8]. They however operate only at a finite frequency where the ENZ occurs. ENZ metamaterials also typically possess broadband hyperbolic and anisotropic elliptic dispersions allowing a rich structure of the supported modes, including low and negative group velocity modes[9]. Such metamaterials can be realised on a macroscopic scale using arrays of aligned plasmonic nanorods so that the mode structure of the metamaterial can be efficiently tuned by changing the nanorod array geometrical parameters. In particular, Au nanorod based metamaterials provide tuneable ENZ behaviour throughout both the visible and near infrared spectral ranges via adjustment of structural parameters. While the centrosymmetric crystal lattice of gold, limits the source of the second-order nonlinearity to the surface nonlinear contributions, the nanostructured geometry of the Au nanorod metamaterial provides a large surface area needed for exploiting the intrinsic surface nonlinearity of gold. Due to the fact that this internal structure of the metamaterial and local fields inside the metamaterial are paramount for the description and understanding of SHG processes, the effective medium approximation for the metamaterial description is not applicable for modelling SHG processes and full vectorial, microscopic considerations should be employed.

In this paper, we study intrinsic second-harmonic generation processes in plasmonic nanorod metamaterial slabs and show its enhancement at multiple resonant wavelengths determined by the mode structure of the metamaterial at both fundamental and second-harmonic frequencies. The enhancement related to phase matching due to ENZ is shown to be negligible in the considered geometry of thin metamaterial slab, with both TE and TM modes providing efficient SHG distributed over the entire UV and VIS regions of the spectrum. SHG enhancement by several orders of magnitude, is observed when a modal overlap between the fundamental and second harmonic frequencies takes place, emulating the so-called double-resonant conditions[4,10,11]. To this end, the perturbative hydrodynamic approach for Au nonlinearity description in a broad spectral range has been developed, and microscopic numerical simulations of the near-field and radiated SHG have been performed in both reflection and transmission. By designing such plasmonic nanorod metamaterials, the enhancement of frequency doubling efficiency can be achieved in a desired wavelength range, in the same way as third-order nonlinearity has been designed[12].

The metamaterial studied here is formed by an assembly of aligned metallic (Au) nanorods (12 nm radius, 300 nm length and 60 nm period of a square lattice) hosted in an alumina matrix (Fig. 1). The optical properties of this uniaxial metamaterial can be inferred from the effective medium theory[14] via the diagonal effective permittivity tensor of the form $\hat{\varepsilon}^{eff} = [\varepsilon_x^{eff}, \varepsilon_y^{eff} = \varepsilon_x^{eff}, \varepsilon_z^{eff}]$ (Fig. 1). A TE-polarized incident light with the electric field normal to the nanorod axis always experiences positive effective permittivity $\varepsilon_{x,y}^{eff}$. In contrast, a TM polarized incident light, having an electric field component along the nanorod axis will be governed by the effective permittivity components $\varepsilon_{x,y}^{eff}$ and $\varepsilon_z^{eff}$. In the former case, the metamaterial behaves as a dielectric over the entire spectral range investigated, while in the latter, the optical response of the metamaterial transitions from an anisotropic dielectric behaviour ($\varepsilon_{x,y}^{eff}$, $\varepsilon_z^{eff} > 0$, elliptic dispersion) to a hyperbolic behaviour ($\varepsilon_{x,y}^{eff} > 0$, $\varepsilon_z^{eff} < 0$, hyperbolic dispersion) passing through the ENZ wavelength range at around a wavelength of 710 nm at which $\mathcal{R}e(\varepsilon_z^{eff}) = 0$ determining the effective plasma frequency of metamaterial[9].

The anisotropic metamaterial slab placed on a glass substrate supports a set of TE and TM modes through the near-infrared, visible, and ultra-violet spectral range (Fig. 2 a,b). Above the light line in air, these modes are the cavity modes of the planar slab of the metamaterial. Below the light line, the modes are accessible only with the illumination through the substrate and are leaky waveguided mode of the metamaterial slab[9]. Above the effective plasma frequency of the metamaterial, both TE and TM modes are typical modes of the anisotropic planar waveguide, while below it the metamaterial supports bulk plasmon polaritons[9], and the TM modes may have positive, vanishingly small or negative group velocity. TE modes always behave as modes of anisotropic dielectric.

The modal structure of the metamaterial slab has a crucial effect on both transmitted and reflected SHG enhancement (Fig. 2 c-f). Considering both linear reflectance and SH generation either for TE (Fig. 2a,c,e) or TM incident polarization (Fig. 2b,d,f), it becomes clear that the slab modes are responsible for the enhancement of the observed SHG. The fundamental light coupling to the modes results in the field enhancement which in turn influences SHG intensity. If the generated SHG light is also coupled to the slab mode, which is possible due to the rich dispersion of modes available in the metamaterial cavity, the efficiency of SHG is further increased. This far-field radiated SHG which is solely related to the electronic nonlinearity of Au nanorods forming the metamaterial is TM-polarised in both reflection and transmission irrespectively of the polarisation of the fundamental light. Both reflected and transmitted SHG have comparable intensities. However, depending on the fundamental wavelength, either TM or TE polarised light excited SHG can be up to 2 orders of magnitude higher. This is a direct consequence of the internal structure of the nanorod metamaterial, providing both field enhancement[15] and large surface area (approximately 7 times larger than for a smooth Au film) and the vectorial nature of the SH process, which favours light coupling to the modes supported by the metamaterial slab. Please note that under the excitation at normal incidence, low SHG intensities are observed for all polarisation configurations due to the symmetry breaking requirements for far-field SHG[16,17] (these requirements are not applicable in the near-field region[18]). In this geometry, nonlinear sources derive from the nanorod extremities. For a smooth Au film under normal incidence illumination, the simulated SHG is vanishingly small, within numerical noise, as expected from symmetry considerations.

In order to benchmark the observed SHG intensities from the metamaterial, the reflected-SHG spectrum excited with TM-polarized fundamental light from a smooth Au surface has been calculated using the same model for Au nonlinearity (Fig. 3). The SHG has a featureless wavelength dependence with a sole resonance in the vicinity of the interband transitions in Au. Compared to the SH power reflected from the nanorod metamaterial slab, the SHG from a smooth film is much lower in the visible and near-IR spectral range (up to 2 orders of magnitude), while higher for short wavelengths in the UV (Fig 3 a). Interestingly, a TM (TE)-polarized fundamental light at 1.16 (1.77) eV incident in a substrate glass at an angle of 48º, efficiently couples to the guided modes to give SH higher (lower) by 1.9 (0.7) times compared to a smooth Au under TM-polarised excitation (Fig. 3b-c). For different angles of incidence, there is redistribution of the peak intensities in the SHG spectra according to the dispersion of the involved modes of the slab, but the maximum enhancement measured with respect to the smooth film remains within a similar intensity range. Moreover, strong radiated SHG is also observed in transmission even for the excitation under total internal reflection (TIR) conditions as the dispersion of the metamaterial slab has the modes at the SH frequencies outside TIR range (Fig. 2e,f).

While for the far-field radiated SHG, the integration over the SHG propagation direction smears out the contribution of individual modes, the analyses of the near-field SHG intensities allows one to identify the role of TM and TE modes of the metamaterial slab (Fig. 4). The modes supported by the metamaterial slab were identified by comparing the dispersion to the analytically calculated mode structure in the framework of the effective medium theory[9] (Fig. 4a,b). A good agreement between the simulated and analytical modes allows to label the observed modes corresponding to the minima in the reflection dispersion (See also Supplementary Information, Fig. S2). For TE polarisation, a set of modes of the anisotropic dielectric slab can be identified (denoted with $TE_n$, where n is the mode index). For TM polarisation, bulk plasmon polaritons are supported by the metamaterial slab (the modes denoted with $TM_{h,n}$) at the frequencies below the effective plasma frequency, while modes at higher frequencies correspond to the modes of a conventional elliptic-dispersion dielectric slab and are denoted as $TM_{e,n}$ (Fig. 4 a,b).

The analysis of the SHG spectra allows one to identify the SHG enhancement at the positions of the modes of the metamaterial slab at either fundamental or SH frequencies, considering that SH emission is always TM-polarized (Fig. 4). The strongest SHG enhancement corresponds however to an overlap of the modes at fundamental and second-harmonic frequencies (Fig. 4 c and d)). The spectral features of the far-field (Figs. 2 c,d) and near-field (Figs. 4 e,f) SH spectra are the same with maxima corresponding to the modes of the metamaterial slab. The metamaterial slab modes identified on the SHG spectral dependencies (red circles for either elliptic or hyperbolic modes at the fundamental frequency and green circles for hyperbolic SHG modes, due to the TM polarisation of the generated SH light), unambiguously show the correspondence of the mode position and the SHG enhancement in the single-resonant (when either the fundamental or SH light is coupled to the mode) and especially double-resonant (when both the fundamental and SH light is coupled to the slab modes) conditions. The double-resonant conditions for TE modes occur primarily for the SHG in the UV spectral range at both incident angles of 30º and 48º with the highest efficiency at 2.3 eV and 1.77 eV. High efficiency can be achieved for TM modes in a broader spectral range (cf. Fig. 4e and 4f) due to the involvement of bulk plasmon-polaritons in the hyperbolic spectral range, e.g. at 1.16 eV excitation, where the hyperbolic $TM_{h,1}$ mode at the fundamental frequency overlaps with the $TM_{e,1}$ mode at the SH frequency. Thus, strong SHG enhancement can be observed at multiple "double-resonant" wavelengths allowing for efficient SH generation at multiple wavelength with the same metamaterial slab.

The field confinement near the nanorods forming the metamaterial, associated with the excitation of different modes was simulated and can be seen from the plotted polarisation distributions: the stronger field enhancement the higher the induced SH polarization. For example, one can observe that the mode $TE_3$ at 2.3 eV (Fig. 5a) yields the highest field confinement, followed by $TE_2$ (Fig. S1a), $TE_4$, $TE_5$, and $TE_1$. In addition to the mode spectral position, an overlap integral between the microscopic nonlinear polarization at second-harmonic and fundamental frequencies is important to observe efficient SHG enhancement[7]. The orthogonal to the surface components of both the nonlinear polarization proportional to $E_\perp^{(\omega)}(\vec{r},\omega)E_\perp^{(\omega)}(\vec{r},\omega)$ and the SH field distributions $E_\perp^{(\omega)}(\vec{r},2\omega)$, together with the associated relative overlap $E_\perp^{(\omega)}(\vec{r},\omega)E_\perp^{(\omega)}(\vec{r},\omega)E_\perp^{(\omega)}(\vec{r},2\omega)$ are plotted for the selected modes and both fundamental light polarizations in Figs. 5 and S1, confirming this requirement. The corresponding values of the relative overlap integrals over the nanorod surface are equal to approximately 0.0003 for $TE_3$ - $TM_{e,3}$ and $TE_2$ -$TM_{e,2}$ modes and 0.04 and 0.01 for $TM_{e,1}$-$TM_{e,3}$ and $TM_{h,3}$-$TM_{e,2}$ modes at 2.3 eV and 1.77 eV fundamental light, respectively. . Thus, the relative overlap integrals are of about one order of magnitude higher for TM compared to TE polarized excitation light. For comparison, for the off resonant excitation at 0.4 eV, the field overlap values are much smaller being $3 \cdot 10^{-5}$ for both TE and TM polarisations, respectively. This microscopic picture highlights the importance of mode selection rules responsible for the SH enhancement in nanorod-based metamaterials. Firstly, a high field enhancement at the fundamental frequency favors the respective modes for achieving strong SH enhancement. Secondly, a strong field enhancement at the SH frequency, along with a multiresonant response, enhances SH even further. Last but not least, the symmetry of the two field distributions is crucial allowing for strongest overlap of the resonant fields.

In conclusion, we have numerically studied SHG from an anisotropic hyperbolic metamaterial in different polarisation configurations and for different angles of incidence. We have adopted a perturbative, nondepleted pump approximation of the SHG description with hydrodynamic nonlinearity of Au nanorod meta-atoms forming the metamaterial and also taken into account the internal structure of the metamaterial composite. The simulations show a strongly enhanced SHG when both fundamental and generated light is coupled to the modes of the metamaterial slab, which is facilitated by the modes' dispersion in the hyperbolic regime. Due to a rich spectrum of the modes supported by the hyperbolic anisotropic slab waveguide, the enhanced SHG can be observed in a broad spectral range. The enhancement related to phase matching due to ENZ is shown to be negligible in the considered geometry of a thin metamaterial slab. By designing the plasmonic nanorod metamaterials, the enhancement of frequency doubling efficiency can be achieved in a desired wavelength range by the design of the mode structure of the metamaterial slab.

**Methods.**

*Model for second-order nonlinearity of Au.* For modelling second-order nonlinear response of metamaterials, we characterized the interaction between the nonlinear sources and the metamaterial at the harmonic frequency $(2\omega)$, via the nonlinear wave equation in the frequency domain $\nabla \wedge \nabla \wedge \boldsymbol{E}^{(2\omega)}(\boldsymbol{r}) - k_0^2 \varepsilon^{(2\omega)}(\boldsymbol{r})\boldsymbol{E}^{(2\omega)}(\boldsymbol{r}) = \mu_0 k_0^2 \boldsymbol{P}^{(2\omega)}(\boldsymbol{r})$, where $\boldsymbol{E}^{(2\omega)}$ is the field at the SH, $k_0$ is the wavevector of light in air, $\varepsilon^{(2\omega)}$ is the permittivity of the nonlinear material at the SH, and $\boldsymbol{P}^{(2\omega)}$ is the nonlinear polarizability generated within the nonlinear (nonmagnetic $\mu_0 =$

1) metallic rods of the metamaterial. Nonlinear polarization in conduction band of noble metals, such as gold and silver, is well described by the hydrodynamic model[19]. The latter holds only for the spectral range below the plasma frequency such that the real part of the permittivity is negative, but both away from interband transitions and Coulomb collisions, occurring respectively at high and low frequencies. Following[20,21], both conduction and bound electron losses can be included in the model, and a complex permittivity at both fundamental and SH frequencies, can be then considered. Let us consider the classical hydrodynamic equation for conduction electrons: $m^* \frac{d\boldsymbol{v}}{dt} + m^*\boldsymbol{v}\cdot\nabla\boldsymbol{v} = -e\boldsymbol{E} - \gamma_0 m^*\boldsymbol{v} - e\mu_0 \boldsymbol{v}\wedge\boldsymbol{H}$, where $\boldsymbol{v}(\boldsymbol{r},t) = \boldsymbol{j}_D(\boldsymbol{r},t)/en(\boldsymbol{r},t)$ is the electron velocity, $\boldsymbol{j}_D(\boldsymbol{r},t)$ and $n(\boldsymbol{r},t)$ being electron current density and electron density, respectively. The equation includes a convective term accounting for the acceleration of charges, a linear term in the charge density, a viscosity term containing the damping frequency $\gamma_0$ of conduction electrons and finally the Lorentz force term. Nevertheless, additional terms such us quantum pressure of electron gas are not taken into account in this model[21]. Given the continuity equation: $\dot{n}(\boldsymbol{r},t) = -1/e(\nabla\cdot\boldsymbol{P})$, the hydrodynamic equation can be written in terms of the polarizability $\boldsymbol{P}$. Thus, adopting a perturbative approach in the undepleted pump approximation, one can move into the frequency domain and write the current density at the SH frequency as a function of the fundamental frequency field by collecting terms up to the second order: $\boldsymbol{P}_S^{(2\omega)} = 4\pi\varepsilon_0\big[\beta \boldsymbol{E}^{(\omega)}\nabla\cdot\chi_D^{(\omega)}\boldsymbol{E}^{(\omega)} + \gamma\nabla(\boldsymbol{E}^{(\omega)}\cdot\boldsymbol{E}^{(\omega)})\big]$, where $\gamma = \frac{\omega_p^2}{\omega^2}\frac{\beta}{4}$ and $\beta = \frac{e}{8\pi m^*\omega^2}\left[\frac{\omega^2}{\omega_p^2}\right]\chi_D^{(\omega)}\left[\frac{(2\omega)^2}{\omega_p^2}\right]\chi_D^{(2\omega)}$. The first term accounting for a dot product between the electric field and its divergence is the quadrupole-like Coulomb term while the second term accounting for the gradient of the square of the electric field is produced by convective and Lorentz forces. These terms have a δ-function-like behaviour at the metal surface, thus we have neglected the bulk contribution to second-order response. It is interesting to note that in the limit of a free electron gas when $\beta = \frac{e}{8\pi m^* \omega^2}$, the above polarizability takes the form reported in Ref.[19] Finally, due to the continuity of the electric displacement across the metal surface, we can write the electric field as a function of the spatial varying complex permittivity $\varepsilon_r$, considering both intraband and interband transitions. We, then, obtain the following tangential and orthogonal terms for the current density at the surface

$$\begin{cases} \widehat{\mathbf{Z}} \cdot \mathbf{j}_S^{(2\omega)} = -2i\omega\varepsilon_0\alpha\chi_D^{(2\omega)}\chi_D^{(\omega)^2}(\varepsilon_r + 3)/4\, \mathbf{E}_{s,\widehat{\mathbf{Z}}}^{(\omega)}\mathbf{E}_{s,\widehat{\mathbf{Z}}}^{(\omega)} \\ \widehat{\mathbf{X}} \cdot \mathbf{j}_S^{(2\omega)} = -2i\omega\varepsilon_0\alpha\chi_D^{(2\omega)}\chi_D^{(\omega)^2}\mathbf{E}_{s,\widehat{\mathbf{X}}}^{(\omega)}\mathbf{E}_{s,\widehat{\mathbf{Z}}}^{(\omega)} \end{cases} \quad (1)$$

where $\alpha = \frac{2e\omega^2}{m^*\omega_p^4}$. Here, it should be noted that the proportionality between the current density and the electric field, contains the susceptibility at the fundamental frequency taken twice and the susceptibility at the SH frequency, as expected for a three-wave interaction. While the tangential term is derived only from the quadrupole-like Coulomb term, the orthogonal term is derived by both surface current density terms, resulting in a tangential term being continuous at the surface in the opposite trend to the orthogonal component. From this consideration and from an additional $(\varepsilon_r + 3)/4$ factor follows that the orthogonal term is dominant, which is in agreement with previously observed results[22,23]. Only the orthogonal term is considered thereafter.

*Numerical modelling of linear response and SHG in a metamaterial slab.* For the numerical implementation of the SHG emitted by the metamaterial we used a frequency-domain Finite Element solver (Comsol Multiphysics), in the framework of the undepleted pump approximation described above. The model uses Floquet boundary conditions for a metamaterial square unit cell to mimic a planar infinite slab. Rounded extremities have been considered for rods to avoid the computational complexity associated with unnecessary mesh. The model has also been tested on Au nanoparticles such as nanospheres and nanoprisms leading to a SH emission in agreement with Ref.[24].

FIGURES

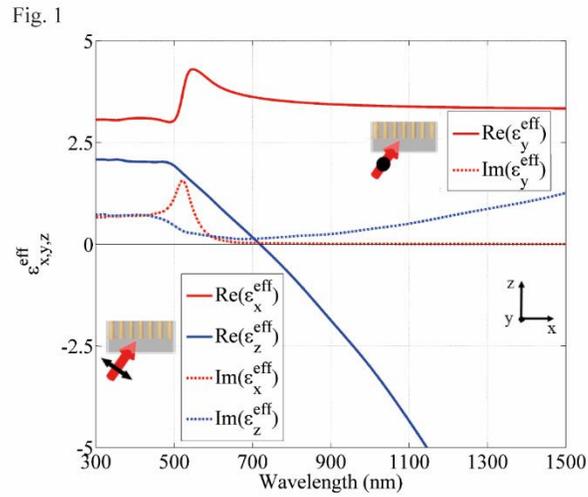

**Figure 1.** Effective permittivity for the anisotropic nanorod metamaterial based on a square array of Au nanorods (12 nm radius, 300 nm length and 60 nm period) embedded in alumina. Au permittivity has been taken from [13] and for alumina n=1.6.

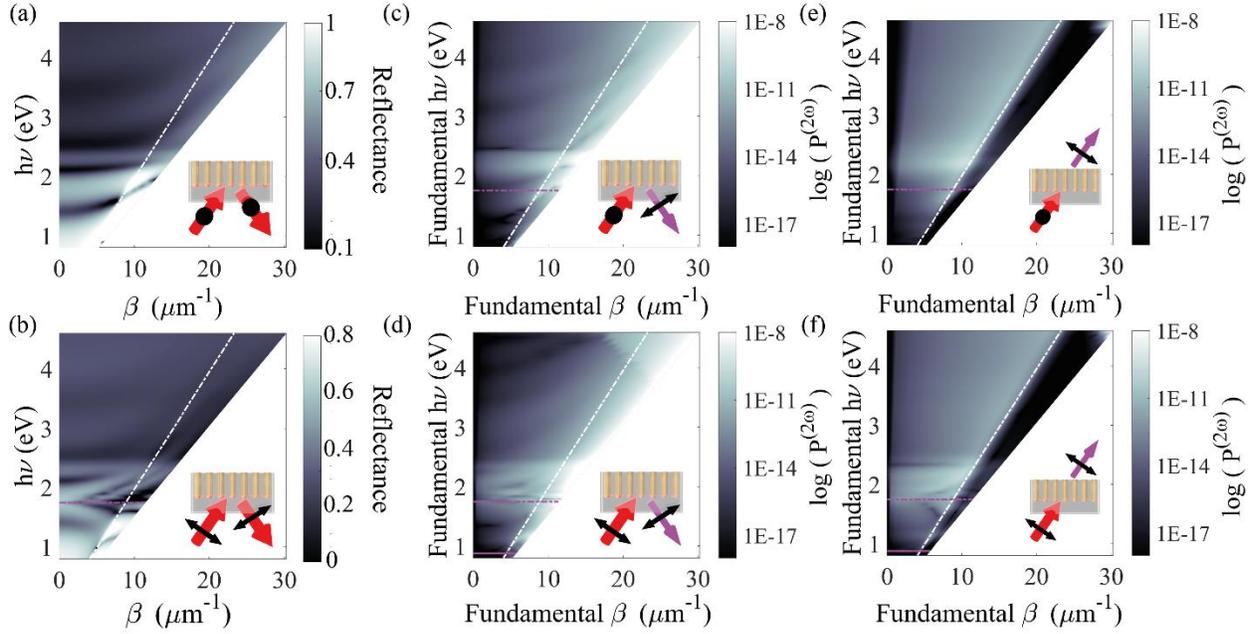

**Figure 2.** (a,b) Reflection dispersion of the metamaterial slab as in Fig. 1 for (a) TE and (b) TM polarisations. (c-f) The spectra of the radiated (far-field) SH intensity for different angles of incidence of the fundamental light: (c,d) reflected and (e,f) transmitted SHG signal for (c,e) TE-polarised and (d,f) TM-polarised fundamental light. SHG is always TM-polarised. The effective plasma frequency of the metamaterial at the fundamental and SH wavelengths is shown with the horizontal dashed and solid lines, respectively. The light line in air is shown with a white dashed line. Please note low SHG intensity near the normal incidence excitation, which is related due to the symmetry consideration of far-field SHG.

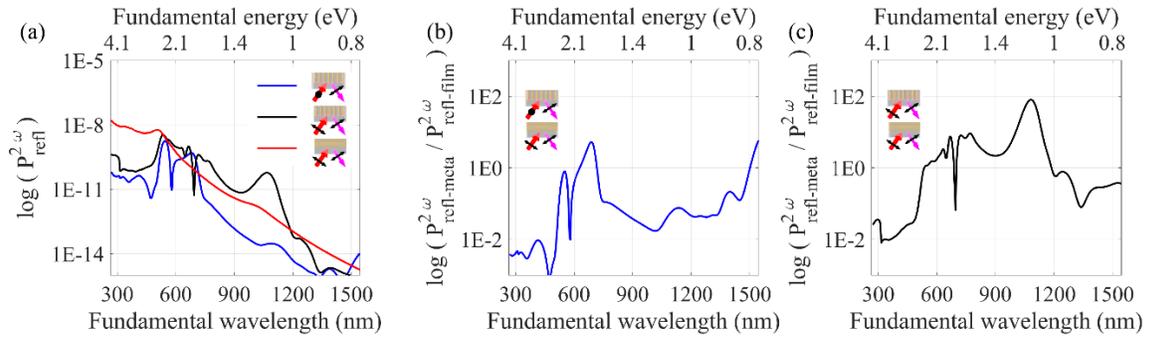

**Figure 3.** (a) Reflected SH spectra from a smooth Au surface (red line) and the nanorod metamaterial (extracted from Fig. 2c,d) for TE-polarised (blue line) and TM-polarised (black line) excitation, respectively. (b,c) The ratio of the SHG intensities reflected from the metamaterial for (b) TM- and (c) TE-polarised excitation and smooth Au surface for TM-polarised excitation. The angle of incidence is 48º in all cases.

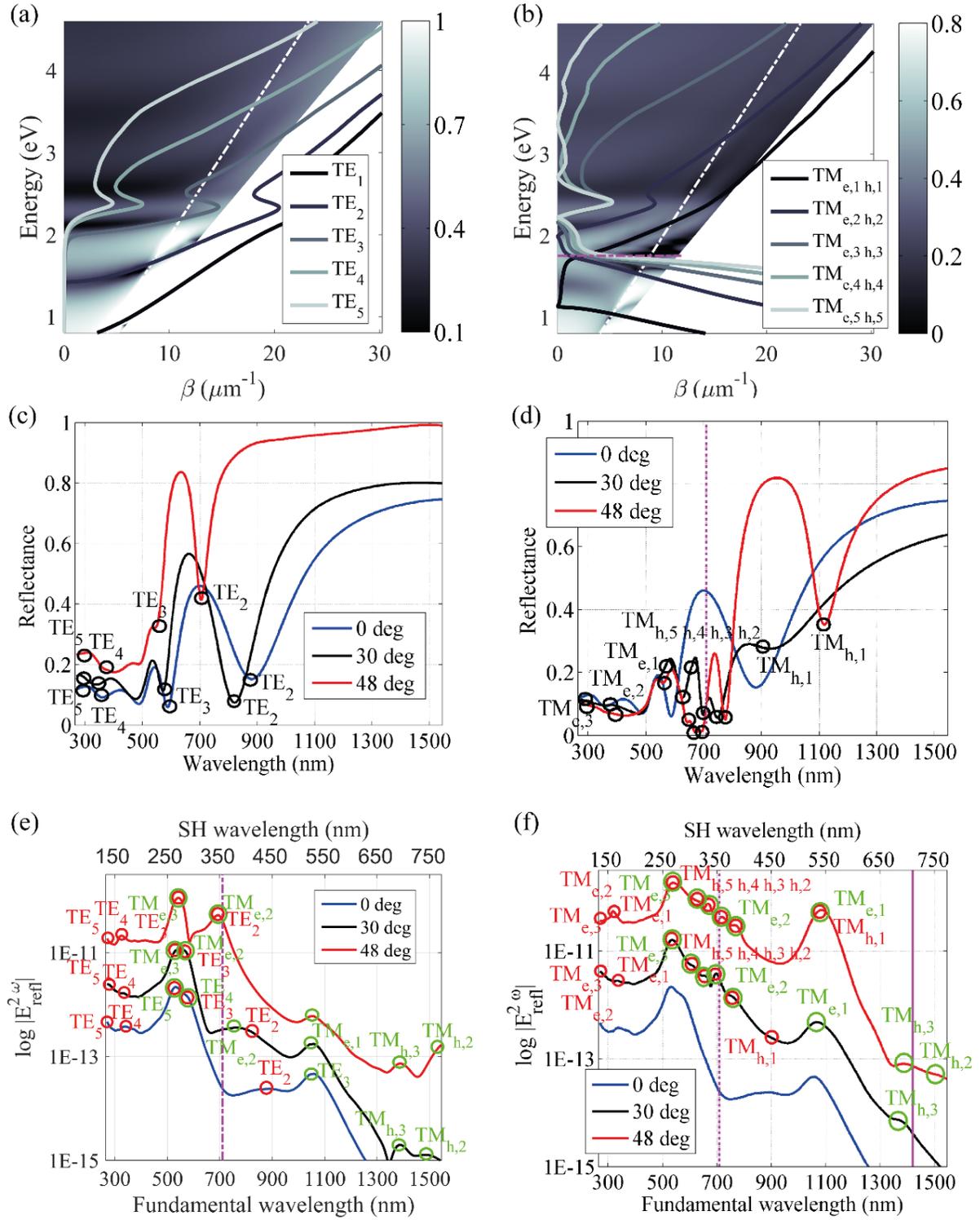

**Figure 4.** (a) TE and (b) TM reflectance dispersion of the metamaterial slab overlaid with the analytically simulated mode structure. The effective bulk plasma frequency of the metamaterial is shown with the horizontal line. (c) TE and (d) TM reflectance spectra for selected wavelengths with mode identification: $TE_n$ modes refer to modes allowed for ordinary polarization, $TM_{e,n}$ and $TM_{h,n}$ modes refer to modes allowed in the elliptical and hyperbolic regimes, respectively, for extraordinary polarization. (e,f) Spectra of the near-field SHG measured 5 nm below the metamaterial in glass substrate for (e) TE and (f) TM polarization of excitation. SH light has TM polarization irrespectively of the fundamental polarization. Red and green circles identify modes at the fundamental and second-harmonic frequencies, respectively. The effective plasma frequency of the metamaterial at the fundamental and SH wavelengths is shown with the horizontal dashed and solid lines, respectively.

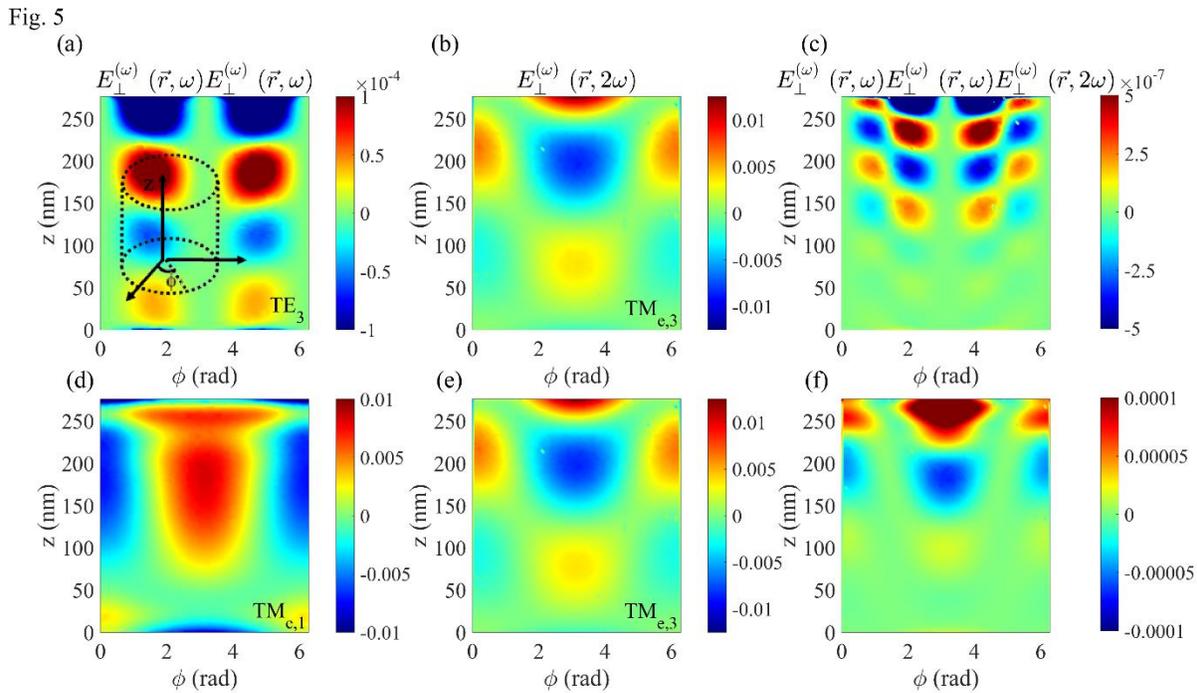

**Figure 5.** (a,d) Orthogonal to the nanorod surface ($P_n$) component of the nonlinear polarization in (a) $TE_3$ and (d) $TM_{e,1}$ modes induced by the TE and TM polarized fundamental light, respectively, at 2.3 eV. (b,e) The electric field distribution of the $TM_{e,3}$ mode at the SH frequency (4.6 eV). (c,f) Relative overlap integral between (c) $TE_3$ and $TM_{e,3}$ modes and (f) $TM_{e,1}$ and $TM_{e,3}$ modes. Fundamental light is incident from the top at the angle of incidence of 48°. The plots are presented around the nanorod circumference.

AUTHOR INFORMATION

**Corresponding Author**

*Giuseppe Marino. giuseppe.marino@kcl.ac.uk.

**Present Addresses**

† Department of Electrical Engineering Physical Electronics, Tel Aviv University, Ramat Aviv 69978, Israel

**Author Contributions**

G.M., G.W and A.Z developed the idea. G.M. made numerical simulations. All authors analysed the results and wrote the manuscript.

ACKNOWLEDGMENTS

This work was supported, in part, by EPSRC (UK), the ERC iPLASMM project (321268), and US Army Research Office (W911NF-12-1-0533). A.Z. acknowledges support from the Royal Society and the Wolfson Foundation. G.W. acknowledges support from the EC FP7 project 304179 (Marie Curie Actions). The data access statement: all data supporting this research are provided in full in the results section in supplementary information.

**SUPPLEMENTARY INFORMATION**

**S1.** Field distribution analysis.

Figure 5 shows the distributions of the selected fundamental and second-harmonic modes in order to illustrate the resonant conditions. The TE mode at fundamental frequency generates the respective TM mode at the second-harmonic frequency. At the fundamental frequency (Fig. 5a) the orthogonal to the surface component of the nonlinear polarization has a clear signature of $TE_3$ with the expected two valleys and a peak. However, at the rod extremities some lightning rods effects are also present. On the other hand, the field distribution at the second-harmonic frequency (Fig. 4b) has two valleys and two peaks, as expected for a $TM_{e,3}$ mode. Significant overlap is obvious. In analogy to Fig. 5a,b, the orthogonal component of both the nonlinear polarization at 1.77 eV and the electromagnetic field at 3.54 eV in Fig. S1a,b, should show the same number of peaks and valleys as the corresponding modes in Ref.[9] One can note that the field at 1.77 eV reports a clear signature of $TE_2$ with the expected one node one valley and one peak. Interestingly, $TE_2$ lies in a spectral range which is away from both interband and intraband transitions, thereby showing sharp modal characteristics. In contrast to this, the map in Figure S1b reports two valleys and a peak, typical of $TM_{e,2}$ mode. Similarly to the case of $TM_{e,3}$, the last valley is less pronounced due to interband transitions and to the arbitrary angle of incidence. Significant relative overlap is obvious in Fig.S1c, which results one order of magnitude lower at the substrate side in comparison with the overlap integral between the modes $TE_3$ and $TM_{e,3}$ (Fig. 5c).

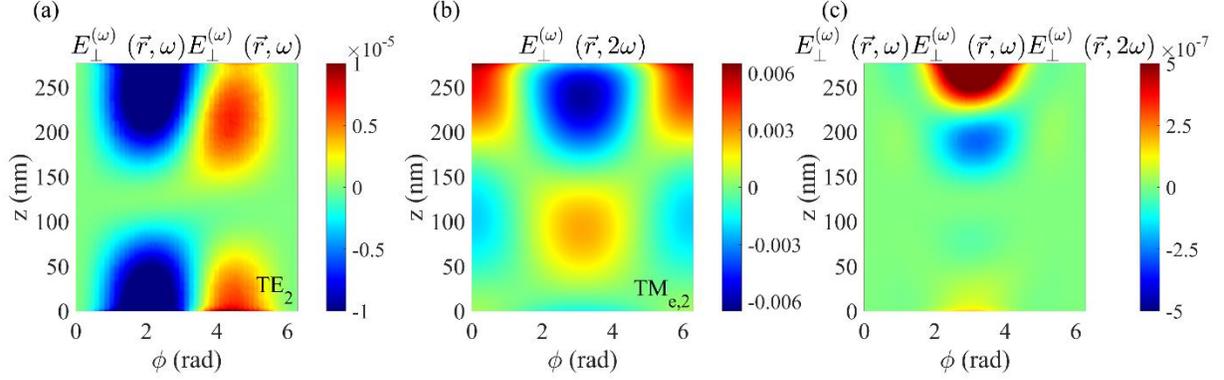

**Figure S1.** (a) Orthogonal to the nanorod surface ($P_n$) component of the nonlinear polarization in $TE_2$ mode induced by the TE polarized fundamental light at 1.77 eV. (b) The electric field distribution of the $TM_{e,2}$ mode at the SH frequency (3.54 eV). (c) Relative overlap integral between $TE_2$ and $TM_{e,2}$ modes. Fundamental light is incident from the top at the angle of incidence of 48°. The plots are presented around the nanorod circumference.

**S2. Analysis of the mode spectrum in the UV**

Due to the high losses in the UV spectral range, especially close to the interband transitions, the identification of the mode in the linear spectra is not straightforward as they become broader[25]. Moreover, superposition of several interband transitions occurring in the UV spectral range results in the characteristic "oscillating" wavelength dependence of the Au permittivity measured in Ref.[13]. In order to deconvolute these sporadic oscillations from the true broad resonances of the metamaterial slab, we simulated the reflectance spectra with a constant value of the complex permittivity typical of gold in the UV ($\varepsilon_r = -1.26 + 5.5i$ at a wavelength of 350 nm) and compared to the reflectance of the metamaterial with the factual permittivity dispersion[13] (Fig. S2). Since the Au permittivity in the UV does not fluctuate far from the value it assumes at 350 nm, we can, then, concentrate our investigation on the mode identification (black line in Fig. S2). Two minima occurring at 400 and 285 nm corresponds to the near field distributions typical of the modes $TM_{e,2}$ and TM$_{e,3}$, respectively. By comparing the field distributions at 400 nm (285 nm) with those at 350 (270 nm), within a FWHM of the respective modes, it is evident that the distributions are the same apart from a small dephasing. Therefore, the oscillations of the blue curve corresponds to the oscillations of the permittivity while the metamaterial slab mode is the same.

Fig. SI 2

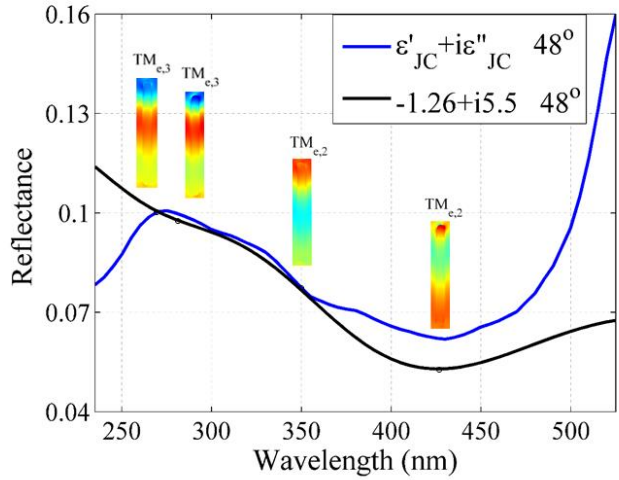

**Figure S2.** Reflectance spectra of the metamaterial for the TM polarized light at 48º angle of incidence simulated for the actual Au permittivity[13] (blue line) and for a fixed dispersionless permittivity (black line). The near-field distributions inside the elementary cell plotted for the dispersionless permittivity at wavelengths of 270 nm, 285 nm, 350 nm, and 400 nm (black circles).